\documentclass[12pt]{JHEP3}
\usepackage{amssymb}
\usepackage{amsmath}
\usepackage{graphicx}
\usepackage{epsfig}
\usepackage{array}
\usepackage{citesort}

\newcommand{\be}{\begin{equation}}
\newcommand{\ee}{\end{equation}}
\newcommand{\bea}{\begin{eqnarray}}
\newcommand{\eea}{\end{eqnarray}}
\newcommand{\ba}{\begin{eqnarray}}
\newcommand{\ea}{\end{eqnarray}}
\newcommand{\bln}{\begin{align}}
\newcommand{\eln}{\end{align}}
\newcommand{\bst}{\begin{split}}
\newcommand{\est}{\end{split}}
\newcommand{\bi}{\begin{itemize}}
\newcommand{\ei}{\end{itemize}}

\newcommand{\ben}{\begin{enumerate}}
\newcommand{\een}{\end{enumerate}}

\def\ov{\over}
\def\le{\left}
\def\ri{\right}
\def\ha{{1\over 2}}

\newcommand{\p}{\partial}

\def\lam{{\lambda}}

\def\Om{{\Omega}}

\def \ga {\gamma}
\def \lam {\lambda}

 \def\Sig{{\Sigma}}
\def\VV{{\cal V}}

\newcommand{\eps}{\varepsilon}
\def\apr{{\alpha'}}

\def\LL{{\cal L}}
\def\TT{{\cal T}}
\def\NN{{\cal N}}

\relax

\title{Velocity dependence of
baryon screening in a hot strongly coupled plasma}

\author{Christiana Athanasiou, Hong Liu and Krishna Rajagopal
\\
\vspace{0.1in}

Center for Theoretical Physics, Massachusetts Institute of Technology,
\\
Cambridge, MA 02139, USA
\vspace{0.1in}

E-mail addresses: {\tt athanasi@mit.edu, hong\_liu@mit.edu, krishna@ctp.mit.edu} }

\abstract{
The $L$-dependence of the static
potential between $N_c$ quarks arranged in a circle
of radius $L$ (a ``baryon") immersed in the hot plasma
of a gauge theory with $N_c$ colors defines a screening length $L_s$.
We use the AdS/CFT correspondence to
compute this screening length for the case of heavy quarks in
the plasma of strongly coupled
$\mathcal{N}=4$ super Yang-Mills  theory moving with velocity
$v$ relative to the baryon.  We find that in the $v\rightarrow 1$ limit,
$L_s \propto (1-v^2)^{1/4}/T$, and find that corrections to this velocity
dependence are small at lower velocities.  This result provides
evidence for the robustness of the analogous behavior
of the screening length defined by the static quark-antiquark pair,
which has been
computed previously and in QCD is relevant to quarkonium
physics in heavy ion collisions.  Our results also show that 
as long as the hot wind is not blowing precisely perpendicular
to the plane of the baryon configuration that we analyze,
the $N_c$ different quarks are not all affected by the wind
velocity to the same degree, with those quarks lying perpendicular
to the wind direction screened most effectively.
}

\preprint{MIT-CTP-3919}

\numberwithin{equation}{section}
\begin{document}

\section{Introduction and Summary}\label{sec:intro}

The simplest example of the AdS/CFT correspondence is provided by the duality between
${\cal N}=4$ supersymmetric Yang-Mills (SYM) theory and classical gravity in
$\textrm{AdS}_5\times {\textrm S}_5$~\cite{AdS/CFT}. ${\cal N}=4$
super Yang-Mills theory is a conformally invariant theory
with two parameters: the rank of the gauge group $N_c$ and the 't~Hooft
coupling $\lambda = g_{\rm YM}^2 N_c$. In the large $N_c$
and large $\lam$ limit, gauge theory problems can be solved using
classical gravity in ${\textrm{AdS}}_5 \times {\textrm S}_5$ geometry. We shall work
in this limit throughout this paper.

In $\NN=4$ SYM theory at zero temperature, the static potential
between a heavy external quark and antiquark separated by
a distance $L^{\rm meson}$ is given in the large $N_c$ and large
$\lam$ limit by~\cite{Rey:1998ik,Maldacena:1998im}
 \be \label{zepo}
 V (L) = -\frac{4\pi^2}{\Gamma(\frac{1}{4})^4}\frac{\sqrt{\lam}}{L^{\rm meson}}\, ,
 \ee
where the $1/L^{\rm meson}$ behavior is required
by conformal invariance.
This potential is obtained by computing
the action of an extremal string world sheet, bounded at
$r\rightarrow\infty$ ($r$ being the fifth dimension of $\textrm{AdS}_5$)
by the world lines of the quark and antiquark and ``hanging down''
from these world lines toward smaller $r$.
At nonzero temperature, the potential
becomes~\cite{Rey:1998bq}
 \bea \label{Awe}
 V(L^{\rm meson},T) \approx & \sqrt{\lam} f (L^{\rm meson}) \qquad & L^{\rm meson}
 < L^{\rm meson}_c \nonumber\\
         \approx & \lam^0 g(L^{\rm meson}) \qquad & L^{\rm meson} > L^{\rm meson}_c\ .
 \eea
In (\ref{Awe}), at $L^{\rm meson}_c = 0.24/T$  there is a change
of dominance between different saddle points and
the slope of the potential changes discontinuously.
When $L^{\rm meson}<L^{\rm meson}_c$, the potential is determined as at
zero temperature by the area
of a string world sheet bounded by the worldlines
of the quark and antiquark, but now the world sheet hangs
down into a different five-dimensional spacetime:
introducing nonzero temperature
in the gauge theory is dual to introducing a black hole
horizon
in the five-dimensional spacetime.
When $L^{\rm meson} \ll L^{\rm meson}_c $,
$f(L^{\rm meson})$ reduces to its zero temperature behavior (\ref{zepo}).
When $L^{\rm meson} > L^{\rm meson}_c$, the potential arises from two disjoint
strings, each separately extending downward from the quark
or antiquark all the way to the black hole horizon. At $L^{\rm meson}\gg L^{\rm meson}_c$,
$g(L^{\rm meson})$ is known and is
determined by  the exchange of the lightest supergravity
mode between the two disjoint strings~\cite{Bak:2007fk}.
It is physically intuitive to interpret $L_c$ as the screening length
$L_s$ of the plasma since at $L_c$ the qualitative behavior of the
potential changes.  Similar criteria are used in the
definition of screening length in QCD~\cite{Karsch:2006sf}, although
in QCD there is no sharply defined length scale
at which screening sets in.   Lattice calculations of the
static potential between a heavy quark
and antiquark in QCD indicate a screening length $L_s \sim 0.5/T$ in hot
QCD with two flavors of light quarks~\cite{Kaczmarek:2005ui} and
$L_s\sim 0.7/T$ in hot QCD with no dynamical
quarks~\cite{Kaczmarek:2004gv}.   The fact that there {\it is} a sharply
defined $L_c$ in (\ref{Awe}) is an artifact of the limit
in which we are working.

In Refs.~\cite{Liu:2006nn,Liu:2006he}, the analysis
of screening was extended to the case of
a quark-antiquark pair moving through the plasma
with velocity $v$.  In that context, it proved convenient
to define a slightly different screening length $L^{\rm meson}_s$,
which is the $L^{\rm meson}$ beyond which no connected
extremal string
world sheet hanging between the quark and antiquark
can be found.  At $v=0$, $L_s^{\rm meson}=0.28/T$~\cite{Rey:1998bq}.
At nonzero $v$, up to small corrections
that have been computed~\cite{Liu:2006nn,Liu:2006he},
\be
 L^{\rm meson}_{s} (v,T) \simeq L^{\rm meson}_s (0,T) (1-v^2)^{1/4}
 \propto \frac{1}{T}(1-v^2)^{1/4}\ .
 \label{lmax}
 \ee
This result, also obtained in Ref.~\cite{Peeters:2006iu} and further
explored in 
Refs.~\cite{Avramis:2006em,Caceres:2006ta,Natsuume:2007vc},
has proved robust
in the sense that it applies in various strongly
coupled plasmas other than
${\cal N}=4$ SYM~\cite{Avramis:2006em,Caceres:2006ta,Natsuume:2007vc}.
(See Refs.~\cite{Dorn:2007zy} for other recent work.)
The velocity dependence of the screening
length (\ref{lmax}) suggests that in
a theory containing dynamical heavy quarks and meson
bound states (which $\NN=4$ SYM does not)
the dissociation
temperature $T_{\rm diss}(v)$, defined as the temperature above which mesons with
a given velocity do not exist,
should scale with velocity as~\cite{Liu:2006nn}
 \be \label{rro}
 T_{\rm diss} (v) \simeq  T_{\rm diss} (v=0) (1-v^2)^{1/4} \ ,
   \ee
since $T_{\rm diss}(v)$ should be the temperature at which
the screening length $L^{\rm meson}_s(v)$ is comparable to the size of
the meson bound state.
The scaling (\ref{rro})
indicates that slower mesons can exist up to
higher temperatures than faster ones.
This result has proved robust in a second sense, in that (\ref{rro})
has also been obtained
by direct analysis of the dispersion relations
of actual mesons in
the plasma~\cite{Mateos:2007vn,Ejaz:2007hg}, 
introduced by adding heavy quarks described
in the gravity dual by a D7-brane whose fluctuations are the mesons~\cite{Karch:2002sh}.
These mesons have a limiting velocity whose temperature dependence
is equivalent to (\ref{rro})~\cite{Ejaz:2007hg}, up to few percent corrections that have been
computed~\cite{Ejaz:2007hg}.

\begin{figure}
  \centering
  \includegraphics*[width=0.9\columnwidth]{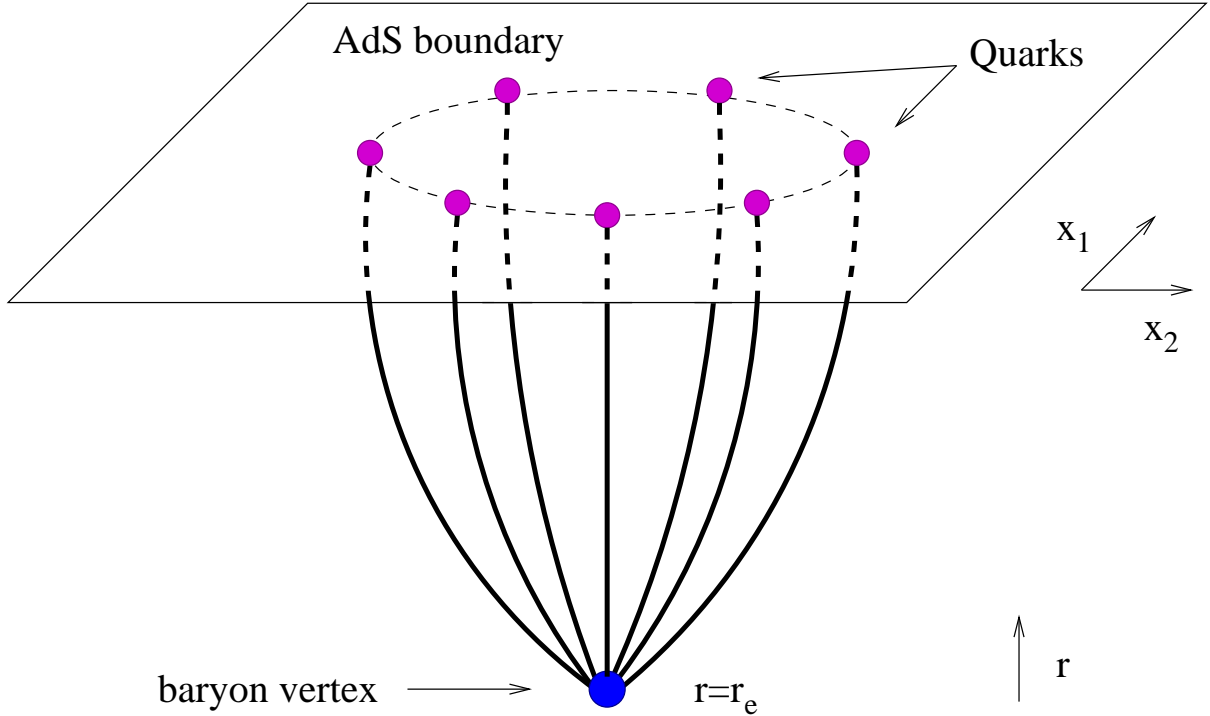}
  \caption{A sketch of a baryon configuration with $N_c$ quarks arranged
  in a circle at the boundary of the AdS space, each connected to a D5-brane located
  at $r=r_e$ by a string.
    }\label{fig:baryon}
\end{figure}

In the present paper, we shall return to the velocity-dependent
screening length and test the robustness of (\ref{lmax}) in yet a
third sense, by analyzing the potential and screening length
defined by a configuration consisting of $N_c$ external quarks
arranged in a circle of radius $L$.\footnote{The baryon static
potential between three static quarks has been computed in QCD
itself using lattice methods at zero
temperature~\cite{LatticeBaryonPotential}, and very recently the
extension of these studies to nonzero temperatures and hence the
study of baryon screening in QCD has been
initiated~\cite{Hubner:2007qh}.} In the gravity dual, there is a
string hanging down from each of these quarks and at nonzero $T$
and large enough $L$, the only extremal configuration of these
string world sheets will be $N_c$ disjoint strings.   In order to
obtain a baryon-like configuration, we introduce a D5-brane into
the gravity dual theory which fills the 5 spatial dimensions of
the S$_5$ and sits at a point in AdS$_5$, and on which $N_c$
strings can 
end~\cite{Witten:1998xy,Gross:1998gk,Brandhuber:1998xy}.\footnote{We
shall only consider the case where all $N_c$ strings are located at the
same point in the S$_5$; it would be interesting to generalize our
analysis to the case where there are different species of quarks
corresponding to strings located at different points in the
S$_5$ which could then end at different points on the 
D5-brane.  We are also neglecting the interactions between the $N_c$ 
string endpoints on the D5-brane.  Such interactions can be
described via the Born-Infeld action for the D5-brane,
and have been analyzed in Refs.~\cite{Imamura:1998gk}
for the case where 
the baryons are BPS objects and the analysis
can be pushed through to completion.  
In our case, in which supersymmetry
is broken by the nonzero temperature and in which the baryon
configurations need not be BPS objects even at zero temperature,
such an analysis certainly presents 
technical challenges and may even be made uncontrolled by
potential higher derivative corrections to the D5-brane
Born-Infeld action. We shall follow
Refs.~\cite{Witten:1998xy,Gross:1998gk,Brandhuber:1998xy} in neglecting
string-string interactions.  
We shall find that the velocity dependence of the screening
length is controlled by the kinematics of the AdS black hole metric
under boosts:  the fact that we find (see below) the same velocity dependence
for the screening length defined by a baryon configuration which
includes a D5-brane as has 
been found previously for that defined by the quark-antiquark potential
whose calculation involves no D5-brane 
suggests, but absent a calculation does not 
demonstrate, that the addition of string-string
interactions on the D5-brane will not modify our conclusions.
} 
This now means that for $L$
less than some $L_s$ we can find configurations as in
Fig.~\ref{fig:baryon}, in which the $N_c$ strings hanging down
from the quarks at the boundary of AdS$_5$ end on the D5-brane.
Following
Ref.~\cite{Brandhuber:1998xy}, we have made the arbitrary choice
of placing the $N_c$ quarks in a circle; pursuing our analysis to
the point of phenomenology would certainly require investigating
more generalized configurations.\footnote{Note that for $N_c=3$,
there is no loss of generality in choosing the $N_c$ quarks to lie
in a single plane, but there is still an infinite space of
distinct possible configurations to consider. Many have been considered
in lattice
investigations~\cite{LatticeBaryonPotential,Hubner:2007qh}.
It would also be worthwhile to extend our analysis to treatments
of baryons themselves, rather than simply calculating the
screening length defined by a particular
(in our case circular) configuration of $N_c$ quarks.
Such an analysis would not directly address the question we pose, namely
the robustness of the velocity dependence of screening, but it would certainly
be a significant step toward actual baryon phenomenology.  The methods
developed in Refs.~\cite{Imamura:1998gk} could provide a good starting point.} 
Our central purpose, however, is to test the
robustness of (\ref{lmax}) in a theoretical context in which the
D5-brane introduces a qualitatively new element.  Note that in
comparing our results for baryon screening to (\ref{lmax}), if we
want to compare numerical prefactors we should compare $L$ to
$L^{\rm meson}/2$, since we have defined $L$ as the radius of the
circle in Fig.~\ref{fig:baryon} rather than its diameter.

The D5-brane plays a role somewhat analogous what has been called a
``baryon-junction" in various phenomenological analyses of baryons
in QCD~\cite{JunctionPhenomenology}.  Baryon junctions in
phenomenological analyses have usually been envisioned as well localized
in (3+1)-dimensions, but this may not be the appropriate way of thinking
of the D5-brane.  The IR/UV relationship that characterizes
the AdS/CFT correspondence~\cite{IRUV}
tells us that smaller values of the fifth-dimension
coordinate $r$ correspond to larger length scales $R^2/r$ in the (3+1)-dimensional
field theory, where $R$ is the curvature of the AdS space.  The D5-brane is
located at $r=r_e$, the lowest point in $r$ of any part of the baryon configuration
in Fig.~{\protect\ref{fig:baryon}}.  It therefore represents the longest
wavelength ``disturbance'' of the (3+1)-dimensional gluon field
(and other ${\cal N}=4$ SYM fields) caused by the presence of the
$N_c$ quarks.  We shall see in Section 3 that in ${\cal N}=4$ SYM
this length scale $R^2/r_e$ is comparable to
$2L$, meaning that the baryon vertex describes a disturbance
of the gluon fields comparable in size to the circle of
external quarks, not 
a baryon junction that is localized in $(3+1)$-dimensions.

The results (\ref{lmax}) and (\ref{rro}) have a simple physical
interpretation which suggests that they could be applicable to a wide
class of theories regardless of specific details.
First, note that since $L_s (0) \sim {1 \ov T}$,  both (\ref{lmax})
and (\ref{rro}) can be interpreted as if in their rest frame the
quark-antiquark dipole experiences a higher effective temperature
$T\sqrt{\ga}$.  Although this is not literally the case in
a weakly coupled theory, in which the dipole will see
a redshifted momentum distribution
of quasiparticles coming at it from some directions and
a blueshifted distribution from others~\cite{Chu:1988wh}, we
give an argument below for how
this interpretation can nevertheless
be sensible. The result (\ref{lmax})
can then be seen as validating the relevance of this interpretation
in a strongly coupled plasma.
The argument is based on the idea that quarkonium propagation and
dissociation
are mainly sensitive to the local energy density of the medium. Now, in the
rest frame of the dipole,
the energy density $\varepsilon$ is
blue shifted by a factor $\sim\ga^2$ and since
$\varepsilon \propto T^4$ in a conformal theory, the result (\ref{lmax}) is
as if quarks feel a higher effective temperature given
by $T\sqrt{\ga}$.
Lattice calculations indicate that the quark-gluon
plasma in QCD is nearly conformal over a range of
temperatures $1.5 T_c < T \lesssim 5 T_c$,
with an energy
density $\varepsilon \approx b T^4$ where $b$ is approximately
constant, at about $80\%$
of the free theory
value~\cite{Karsch:2000ps}. So it does not seem far-fetched to
imagine that (\ref{lmax}) could apply to QCD.
We should also
note that AdS/CFT calculations in other strongly coupled gauge
theories with a gravity description are consistent with the
interpretation above~\cite{Caceres:2006ta,Ejaz:2007hg} and that
for near conformal theories the deviation from
conformal theory behavior appears to be small~\cite{Caceres:2006ta}.
If a velocity scaling like (\ref{lmax}) and (\ref{rro}) holds for
QCD, it can potentially have important implications for quarkonium
suppression in heavy ion collisions~\cite{Liu:2006nn,Ejaz:2007hg},
in particular suggesting that in a heavy ion collision at RHIC (or LHC)
which does not achieve a high enough temperature to
dissociate $J/\Psi$ (or $\Upsilon$) mesons at rest, the production of
these quarkonium mesons with transverse
momentum above some threshold may nevertheless
be suppressed~\cite{Liu:2006nn}.\footnote{These phenomenological
implications rest as much on the analysis of 
the velocity dependence of screening, introduced in~\cite{Liu:2006nn} for
mesons and generalized here to baryons, as they do on the construction
of the dispersion relations of the mesons themselves as in~\cite{Ejaz:2007hg}.
The dispersion relations extend to arbitrarily large wave vectors: there
is a limiting velocity but, for $\lambda\rightarrow\infty$, 
no limiting momentum.  At finite $\lambda$, the mesons have
nonzero widths~\cite{Ejaz:2007hg}; if these widths grow
with increasing meson momentum, this could serve to limit
the meson momenta also~\cite{MyersPrivateCommunication}.
In the absence of widths, as for $\lambda\rightarrow\infty$, inferences
about meson production rely upon the observation that the potential between
a quark and antiquark moving with high enough velocity is screened, making it
unlikely that they will bind into a meson even though a slowly moving meson
state with the same momentum as the quark and antiquark pair does 
exist~\cite{Ejaz:2007hg}.  Thus, if we are to use a baryon analysis to test
the robustness of phenomenological conclusions drawn in the meson
sector, a key point to test is the velocity dependence of screening. Doing
so, as in this paper, does not require analysis of the dissociation of actual baryons.}
Our results suggest that if baryons containing three charm
quarks are ever studied in heavy ion collision experiments,
the suppression of their production could be similarly
dependent on transverse momentum.

In Section~\ref{sec:gen} we shall set up a general formalism for finding
baryon configurations of heavy external quarks in supergravity, with
the $N_c$ quarks arranged arbitrarily.  In 
Section~\ref{sec:main} we shall apply this general formalism to the configuration
depicted in Fig.~\ref{fig:baryon}, allowing us to define
a screening length $L_s$. In Section~\ref{sec:perpend} we evaluate $L_s(v,T)$ for the
case where the baryon configuration is moving through the plasma in
a direction perpendicular to the plane defined by the circle of
quarks. (Equivalently, the ``baryon" feels a plasma wind blowing in
a direction perpendicular to its plane.)
Static configurations are found by extremizing
the total baryon action coming from both the strings and the D5
brane.  We find static configurations only for $L<L_s(v,T)$ with
$L_s(0,T)=0.094/T$ as in~\cite{Brandhuber:1998xy},
comparable to $\frac{1}{2}L^{\rm meson}(0,T)$
above, and with
\be
L_s(v,T)= \frac{0.083}{T}(1-v^2)^{1/4}
\label{PerpendicularResult}
\ee
in the $v\rightarrow 1$ limit.  In this limit, we obtain (\ref{PerpendicularResult}) analytically.
We find numerically that $L_s(v,T) T /(1-v^2)^{1/4}$ varies monotonically and smoothly
from $0.094$ at $v=0$ to $0.083$ at $v\rightarrow 1$, making
\be
\label{SimpleScaling}
L_s(v,T)\simeq L_s(0,T) (1-v^2)^{1/4}
\ee
a good approximation.
In Section \ref{sec:parallel} we do a
similar numerical calculation for the case where the wind velocity
is parallel to the baryon's plane.
At high velocities we find a result like (\ref{PerpendicularResult}) except
that the proportionality constant is different for different quarks/strings,
depending weakly on the angle
between the wind velocity and the string.  $L_s$ is smallest for the quarks
whose strings are oriented perpendicular to the wind, even though
in the configuration that we analyze these quarks are also closest
to the D5-brane. This indicates that as $v$ increases the medium 
is most effective at screening the potential felt by these quarks.

\section{General baryon configurations} \label{sec:gen}

We wish to analyze a baryon configuration of $N_c$ heavy external
quarks in the $\mathcal{N}=4$ SYM plasma at nonzero temperature.
The baryon
construction in supergravity involves $N_c$ fundamental strings with
the same orientation, beginning at the heavy quarks on the AdS
boundary and ending on the baryon vertex in the interior of
$\textrm{AdS}_5$, which is a D5 brane wrapped on the $\textrm{S}_5$
\cite{Witten:1998xy}.  In this Section, we shall allow the $N_c$ quarks
to be placed at arbitrary positions in the $(x_1,x_2,x_3)$-space 
at the boundary of  AdS.
Note that the $\mathcal{N}=4$ SYM plasma
contains no particles in the fundamental representation, so the
quarks we study here are external.   

The gravity theory dual to $\NN=4$ SYM theory at
nonzero temperature is the AdS black hole times a 
five-dimensional sphere, with the metric
 \be
 ds^2 =  - f(r)  dt^2 +  \frac{r^2}{ R^2} d\vec x^2 +  \frac{dr^2}{f(r)} + R^2 d \Om_5^2,
\label{ZeroVelocityMetric}
 \ee
 where
 \be
f(r)=\frac{r^2}{R^2}\left(1-\frac{r_0^4}{r^4}\right).
\label{fdefn}
 \ee
Here, $d \Om_5^2$ is the metric for a unit $S_5$, $R$ is the curvature radius of the
AdS metric, $r$ is the coordinate of the fifth dimension of
AdS$_5$ and $r_0$ is the position of the black hole horizon. 
The
temperature of the gauge theory is given by the Hawking
temperature of the black hole, $T=r_0/(\pi R^2)$. And, the gauge
theory parameters $N_c$ and $\lambda$ are given by
$\sqrt{\lambda}=R^2/\alpha'$ and $\lambda/N_c = g^2_{\rm YM} =
4\pi g_s$ where $1/(2\pi\alpha')$ is the string tension and $g_s$
is the string coupling constant.  (So, large $N_c$ and $\lambda$
correspond to large string tension and weak string coupling and
thus justify the classical gravity treatment.)

We shall always work in the rest frame of the baryon configuration.
This means that in order to describe $N_c$ quarks moving through
the plasma with velocity $v$, say in the $x_3$-direction,
we must boost the metric (\ref{ZeroVelocityMetric}) such
that it describes a $\NN=4$ SYM plasma moving with a wind velocity
$v$ in the negative $x_3$-direction. We obtain 
\begin{equation} \label{eq:1}
ds^2=-A dt^2 +
2B\,dt\,dx_3+C\,dx^2_3+\frac{r^2}{R^2}\left(dx^2_1+dx^2_2\right)+\frac{1}{f(r)}dr^2+R^2
d\Omega^2_5\,,
\end{equation}
where
\begin{equation} \label{eq:2}
A=\frac{r^2}{R^2}\left(1-\frac{r^4_1}{r^4}\right), \qquad
B=\frac{r^2_1 r^2_2}{r^2 R^2}, \qquad
C=\frac{r^2}{R^2}\left(1+\frac{r^4_2}{r^4}\right),
\end{equation}
with
\begin{equation} \label{eq:3}
r^4_1=r^4_0\cosh^2\eta, \quad \textrm{and} \quad
r^4_2=r^4_0\sinh^2\eta .
\end{equation}
We have defined the wind rapidity $\eta$ via
$v=- \tanh\eta$.
Although in Section 3 we shall specialize to circular baryon configurations
as illustrated in Fig.~\ref{fig:baryon}, in this 
Section we describe the construction
of a baryon configuration with $N_c$ heavy 
external quarks placed at arbitrary locations $(x_1,x_2,x_3)$
at $r\rightarrow\infty$ in the boosted AdS metric (\ref{eq:1}).

The construction 
in this Section can easily be generalized
to baryon configurations a large class of gauge theories
at nonzero temperature, including $\NN=4$ SYM
as one example.  Consider any gauge theory that is
dual in the large $N_c$ and strong
coupling limit to Type IIB string theory in the supergravity
approximation in a generic string frame metric that can be written in the form
\be \label{GenericMetric}
 ds^2 =   g_{\mu \nu} (r) dx^\mu d x^\nu +  {dr^2 \ov f(r)} + e^{2
 \psi (r)} ds_5^2\ ,
 \ee
with the possibility of a nontrivial dilaton $\phi(r)$. 
As before, $x^\mu = (t, \vec x) = (t, x_1, x_2, x_3)$ describe the
Yang-Mills theory coordinates (the boundary coordinates). 
Here, $ds^2_5$ is the metric of some
five-dimensional compact manifold $M_5$  that may not be $S_5$.
A specific choice of gauge theory will correspond to specific
choices of $\phi(r)$ and the various metric functions appearing in (\ref{GenericMetric}).
The metric (\ref{GenericMetric}) is not even the most general that we could analyze,
since for example we have not allowed the metric functions in (\ref{GenericMetric}) 
to depend on
the coordinates of the internal manifold $M_5$ and since we
have chosen the $r$-dependence of the $M_5$-metric to be
a common factor  $\exp(2\psi(r))$, not some more complicated
structure.  Such complications do not add qualitatively new features to  
the analysis of baryon configurations in a metric
of the form (\ref{GenericMetric}). Our construction
of baryon configurations below starting from the metric (\ref{GenericMetric})
could be applied to gauge theories known to have
dual gravity descriptions some of which are  conformal and some not, without or with
nonzero $R$-charge density, with $\NN=4$ supersymmetry or
to certain theories with only $\NN=2$ or $\NN=1$ supersymmetry, 
at nonzero or zero temperature, with or without a wind velocity.  In our explicit 
definition of and
calculation of the screening length $L_s$ in Section 3, we shall
return to the special case (\ref{eq:1}) of hot $\NN=4$ SYM theory with a wind
velocity.

A baryon configuration in
the supergravity metric (\ref{GenericMetric})  
involves $N_c$ fundamental strings
beginning at the external heavy quarks on the boundary (which we
will take to be at $r=\infty$) and ending on the baryon vertex in
the interior, which is a D5 brane wrapped on the compact manifold
$M_5$~\cite{Witten:1998xy}. 
We denote the positions in $\vec x$-space
where we place the external quarks by
$\vec q^{(a)}$, with $a=1, \cdots N_c$, and 
we take all the quarks to 
sit at the same point in the compact manifold $M_5$.
We shall describe how to determine the location of the 
D5-brane below. After so doing, we shall shift the origin
of the $\vec x$ coordinates such that the D5-brane sits
at the origin, at  $\vec x_e=0$. We denote its position in the fifth
dimension by $r=r_e$. 
The total action of the system is then given by
  \be \label{eq:7_2}
  S_{\rm total} =
  \sum_{a=1}^{N_c}  S_{\rm string}^{(a)} + S_{\rm D5}\ ,\
 \ee
where $S_{\rm string}^{(a)}$  denotes the action of the
fundamental string connecting the $a$-th quark with the D5-brane.
Denoting the string worldsheet coordinates $(\tau,\sigma)$, we can
choose
\begin{equation} \label{eq:5}
\tau = t, \qquad \sigma=r, \qquad x_i =x_i (\sigma) \ ,
\end{equation}
meaning that the shape of the $a$'th  string worldsheet is
described by functions $x_i^{(a)}(r)$ that extend
from $\vec x^{(a)}(r_e)=\vec x_e$ to $\vec x^{(a)}(\infty)=\vec q^{(a)}$.
The Nambu-Goto action of one string can then be written as 
%
\begin{equation} \label{eq:6}
S_{\mathrm{string}}=\frac{\mathcal{T}}{2\pi
\alpha'}\int_{r_e}^{\infty} dr
 \sqrt{-{g_{00} \ov f} + \left(g_{0i} g_{0j}- g_{00} g_{ij}\right) x_i' x_j'}
 \equiv \frac{\mathcal{T}}{2\pi \alpha'} \int dr \:
\mathcal{L}_{\mathrm{string}},
\end{equation}
where $\mathcal{T}$ is the total time 
and where $x'_i\equiv \p_r x_i$.
 The action for the five-brane can be written as
 \be \label{d5A}
S_{\rm D5} = {\VV (r_e) \TT V_5 \ov (2 \pi)^5 \apr^3 } , \qquad
\VV (r) = \sqrt{-g_{00}} e^{-\phi + 5 \psi},
 \ee
where $V_5$ is the volume of the compact manifold $M_5$ and
$\VV(r_e)$ can be considered to be the gravitational potential for
the D5-brane located at $r=r_e$.  

In order to find a static baryon configuration, we must extremize
$S_{\rm total}$, first with respect to the functions $x_i^{(a)}(r)$ that
describe the trajectories of each of the $N_c$ strings and second
with respect to $\vec x_e$ and $r_e$, the location of the D5-brane.
Because $S_{\rm total}$ does
not depend on the $x_i^{(a)}(r)$ explicitly, 
the variation with respect to $x_i^{(a)}(r)$ leads to Euler-Lagrange
equations that have a first integral
\begin{equation} \label{eq:8}
{\p \LL _{\rm string}^{(a)} \ov \p x_i'^{(a)}} = {\left(g_{0i} g_{0j}- g_{00} g_{ij}\right)
 x_j'^{(a)} \ov  \LL _{\rm string}^{(a)}} = \textrm{const}. \equiv K_i^{(a)} \ ,
\end{equation}
where we have denoted the integration constants by
$K_i^{(a)}$. 
Next, we extremize the action with respect to variations
in the position of the D5-brane, understanding that as 
we vary its position we adjust the string trajectories as
required by their Euler-Lagrange equations.  
Extremizing the action with respect to the location of the
D5-brane in $\vec x$-space yields equations which
receive one contribution from the boundary term at the D5-brane
at $r=r_e$  in the variation of each of the $x_i^{(a)}(r)$, equations which
take the form  
  \be \label{Bss}
\sum_{a} K_i^{(a)}  = 0 \ .
 \ee
(What arises from the variation are the $K_i^{(a)}$ evaluated
at $r=r_e$, but the $K_i^{(a)}$ are by construction $r$-independent.)
The constraint (\ref{Bss}) is a force balance condition, encoding the
requirement that in a static baryon configuration the net force exerted by the
$N_c$ strings on the D5-brane
in the $x_i$ directions, with $i=1$, 2 and 3, must vanish.
Extremizing $S_{\rm total}$ with respect to $r_e$ yields
the $r$-direction force balance condition which we can write as
 \be \label{vre}
  \sum_{a=1}^{N_c} H^{(a)} \biggr|_{r_e} =   \Sig, \qquad
 \ee
where
 \be \label{deH}
 H^{(a)} \equiv
 \LL^{(a)} -  x_i'^{(a)} {\p \LL^{(a)} \ov \p x_i'^{(a)}}
 = {- g_{00} \ov f(r)  \LL^{a}_{\rm string}}
 \ee
is the ``upward'' (i.e. in the positive $r$-direction) force on the D5-brane
from the $a$'th string, meaning that the left-hand side of (\ref{vre})
is the upward force due to all the strings, and where 
\be \label{deSigma}
\Sig \equiv {2 \pi \apr \ov \TT} {\p S_{\rm D5} \ov \p
   r_e}  = {V_5 \ov (2 \pi)^4 \apr^2} {\p \VV(r_e) \ov \p r_e} \ 
\ee
is the downward gravitational force on the D5-brane, given its placement
at $r=r_e$ in the curved spacetime (\ref{GenericMetric}).
Including the contribution to the energy from the
interaction among the $N_c$ string endpoints on
the D5-brane (which has been calculated in
simpler settings than ours~\cite{Imamura:1998gk}) would affect 
our calculation only by
modifying this downward force somewhat.

Eqs. (\ref{eq:8}), (\ref{Bss}) and (\ref{vre}) determine the
shape of the string trajectories and the location of the D5-brane,
which is to say that they determine the baryon configuration for
a given choice of the positions of the quarks $\vec q^{(a)}$.
Used in this way, one would integrate the first order equations 
(\ref{eq:8}), using the  
boundary conditions $\vec x^{(a)}(\infty) = \vec q^{(a)}$ 
to determine the integration constants $\vec
K^{(a)} (\vec q^{(a)},\vec x_e,  r_e)$ for a given
choice of $\vec x_e$ and $r_e$. Eqs. (\ref{Bss})
and  (\ref{vre}) can then be
used to determine $\vec x_e$ and $r_e$. 
Not all choices of $\vec q^{(a)}$ will
yield a static baryon configuration.
For a given quark
distribution at the boundary, the question of whether equations
(\ref{eq:8}), (\ref{Bss}) and (\ref{vre}) have solutions is a dynamical question
depending on the specific metric under consideration.
We shall see specific examples of how this plays out in Section 3.

Alternatively, a baryon configuration can be specified by starting
with a set of $\vec K^{(a)}$ satisfying (\ref{Bss}), solving for
$r_e$ using~(\ref{vre}), and 
integrating Eqs.~(\ref{eq:8}) outward from $r=r_e$ to the
boundary at $r=\infty$, only then learning the quark
positions $\vec q^{(a)}$ in the gauge theory. 
Instead of specifying
$\vec K^{(a)}$, one can equivalently 
specify $\vec s^{(a)}\equiv\p_r\vec x^{(a)} (r)|_{r=r_e}$.

Whether we think of specifying conditions at $r=r_e$
and integrating inwards or specifying conditions at the D5-brane,
since we are considering the $N_c \to \infty$ limit
it is often more convenient to introduce
the density of quarks and strings instead of discrete position variables.
At the boundary, the quark configuration can be specified by a
density of quarks $ \rho (\vec q )$, which can be normalized as
 \be
 \int d^3 \vec q
 \, \rho (\vec q) = 1 \ .
 \ee
 We can then rewrite (\ref{Bss}) as 
\be
  \int d^3 \vec q \, \rho(\vec q)  \, \vec K (\vec q) = 0 \ .
 \label{Bss2}
 \ee
However, (\ref{vre}) cannot immediately be written in terms
of $\rho(\vec q)$ because the quantities in (\ref{vre}) are evaluated
at $r=r_e$, and unlike the $K$'s occurring in (\ref{Bss}) are not $r$-independent.
So, we must use the string trajectories themselves to relate the density
of quarks at $r=\infty$ to a density of strings at $r=r_e$, as follows.
For any given $r_e$ and $\vec x_e$,
a solution $\vec x(r)$ to Eqs.~(\ref{eq:8}) describes a single string trajectory
which connects a particular point $\vec q$ at $r=\infty$ to
the D5-brane
at $\vec x(r_e)=\vec x_e$.    The string connects to  the D5-brane
with a particular value of the ``angle''
$\vec s = \p_r \vec x (r)|_{r=r_e}$.
So, the set of string trajectories $\vec x(r)$ with all possible choices of $\vec q$ 
determine a mapping from $\vec q$ onto $\vec s$, where the $\vec q$\,'s
specify the location of quarks at infinity and the $\vec s$\,'s specify strings
at the D5-brane.
Since the mapping corresponds to Hamiltonian ``time'' evolution
(with $r$ playing the role of time) Liouville's theorem tells us that
a given $\rho(\vec q)$ maps onto a 
$\rho_{V} (\vec s)$ that specifies the density of strings
hitting the D5-brane as a function of angle given by
\be
\rho_V(\vec s) = \rho(\vec q) \left| \frac{ \partial\left(q_1,q_2,q_3 \right) }
{\partial\left( s_1,s_2,s_3\right)}\right|\ .
\ee
In evaluating the Jacobian determinant,
the $\vec q$'s should be considered to be functions of the
$\vec s$'s, with the function being
the mapping defined by the string trajectories $\vec x(r)$. 
If the solutions $\vec x(r)$ are nontrivial
curved trajectories, then the relation between
$\rho(\vec q)$ and $\rho_V(\vec s)$ will be nontrivial.  
Eqs.~(\ref{Bss}) and (\ref{vre}) can now be recast 
in terms of 
 $\rho_V (\vec s)$, namely\footnote{Note that in the
continuous limit,
  $$
 {1 \ov N} \sum_a (\cdots)  \to \int d^3 \vec s \,
 \rho_V (\vec s)  (\cdots) = \int d^3 \vec q \,
 \rho (\vec q)
 (\cdots) \ .
 $$}
 \be
  \int d^3 \vec s \, \rho_V (\vec s)  \, \vec K (\vec s) = 0 
\label{Bss3}
 \ee
 and
 \be
 \int d^3 \vec s \, \rho_V (\vec s)  \, H (\vec s)  = {\Sig \ov N_c} \ .
 \label{vre2}
 \ee
Note that $\vec K (\vec s)$ is obtained by evaluating the left
hand side of (\ref{eq:8}) at $r=r_e$, while $H(\vec s)$ is obtained
by evaluating equation (\ref{deH}) at $r=r_e$.  

We close this Section with a description of one way in which
the formalism that we have developed can be used.  Suppose
that we wish to describe a baryon configuration in which the quarks
all lie on some closed two dimensional surface in $\vec x$-space.
For a given $r_e$, we can then use (\ref{Bss})  in the form (\ref{Bss2}) to determine
the density of quarks along the surface required for any choice of 
$\vec x_e$ located inside the surface.  Or, if the density of quarks along
the surface has been specified, we can use (\ref{Bss}) to determine
$\vec x_e$ for a given $r_e$.  We then repeat this exercise for all values
of $r_e$ until we find an $r_e$ that satisfies (\ref{vre}) in the form (\ref{vre2}).

In next Section we apply (\ref{eq:8}), (\ref{Bss}) and (\ref{vre}) to 
particular
baryon configurations in a $\NN=4$ SYM plasma moving
with a nonzero wind velocity.

\section{Velocity dependence of baryon screening in $\NN=4$ SYM
theory}\label{sec:main}

We now refocus on baryon configurations at rest in
the plasma of  $\NN=4$ SYM theory with temperature $T$
moving with a wind velocity $v=-\tanh\eta$ in the $x_3$ direction. The gravity
dual of this hot plasma wind is
described by the metric (\ref{eq:1}).  
Following Ref.~\cite{Brandhuber:1998xy}, we shall analyze
baryon configurations in which the $N_c$ quarks all lie in
a single plane.  In Section~\ref{sec:perpend} we take the
quarks to be uniformly distributed along a circle in the $(x_1,x_2)$-plane,
perpendicular to the direction of the wind.  In Section~\ref{sec:parallel}
we analyze a configuration in which the quarks lie in the $(x_1,x_3)$-plane,
parallel to the direction of the wind. We expect that the two configurations
we shall study are sufficient to illustrate the generic aspects of the
velocity dependence of baryon screening in $\NN=4$ SYM theory.

\subsection{Wind perpendicular to the baryon
configuration}\label{sec:perpend}

In this subsection we consider a baryon configuration lying in the
$(x_1,x_2)$-plane (i.e. $x_3 =0$) perpendicular to the wind direction. For
simplicity, we arrange the $N_c$ external quarks uniformly
around a circle of radius $L$ as in~\cite{Brandhuber:1998xy}, see
Fig.~\ref{fig:baryon}. This is a simple example within which
we can illustrate many aspects of the general formalism
of Section~\ref{sec:gen} for constructing baryon configurations,
and define and study the velocity dependence of the screening length.

With the quarks arranged uniformly around a circle, it is clear by
symmetry that the D5-brane must sit at the center of the circle,
which we shall take to be at the origin: $\vec x_e=0$.   
Because of the rotational symmetry of the circular configuration and
of the background geometry~(\ref{eq:1}),
each of
the $N_c$ strings in Fig.~\ref{fig:baryon} is equivalent.  They all
sit at $x_3=0$, and each can be described by a single function 
$x(r)$, where
$x\equiv \sqrt{x_1^2+x_2^2}$ extends from $x=0$ and $r=r_e$, at
the D5-brane, to $x=L$, at the boundary of AdS$_5$.   With the 
 D5-brane  at $\vec x_e=0$ at the center of the circle, it
is clear that the forces in the $\vec x$ directions exerted by
the strings on the D5-brane cancel, meaning that Eqs.~(\ref{Bss}) are automatically
satisfied.  The D5-brane sits at some $r=r_e$, which we shall determine
for a given $L$ using (\ref{vre}).  So, $x(r_e)=0$
and $x(\infty)=L$. 
Applying
equations (\ref{eq:6}) and (\ref{d5A}) to (\ref{eq:1}), we find
that in this case
 \be \label{eq:7}
  \LL_{\rm string} = \sqrt{A\left(\frac{(x')^2\, r^2}{R^2}+\frac{1}{f(r)}\right)},
\ee
and
\be \label{eq:S_D5}
S_{\mathrm{D5}}
=\frac{N_c \mathcal{T} R \sqrt{A(r_e)}}{8\pi \alpha'}\ , 
 \ee
where $f(r)$ and $A(r)$ were given in Eqs.~(\ref{fdefn}), (\ref{eq:2}) and (\ref{eq:3}).
The equation (\ref{eq:8}) that determines the shape of the string trajectory $x(r)$
becomes
 \begin{equation} \label{eq:81}
\frac{A \,r^2 x'}{R^2 \LL_{\rm string}} = K,
\qquad
\end{equation}
where by symmetry there is only a single integration constant $K$ for all the
strings.  The $r$-direction force balance 
condition (\ref{vre}), namely the condition that the upward force on the D5-brane 
exerted by the $N_c$ strings balances the downward force of gravity, becomes
 \begin{equation} \label{eq:91}
\frac{A}{ f \LL_{\rm string}} \biggr|_{r_e}= \frac{1+\rho^4
\cosh^2\eta}{4 \sqrt{1-\rho^4 \cosh^2\eta}} \equiv
\Sigma(\rho,\eta),
\end{equation}
 where we have defined
 \be \rho\equiv \frac{r_0}{r_e} = \frac{\pi R^2 T}{r_e} .
 \ee
We must solve (\ref{eq:81}) and (\ref{eq:91}) simultaneously,
in so doing obtaining both the position of the D5-brane $r_e$ and
the shape of the strings $x(r)$ corresponding to a static baryon
configuration with size $L$.

\begin{figure}
  \centering
  \includegraphics*[width=0.7\columnwidth]{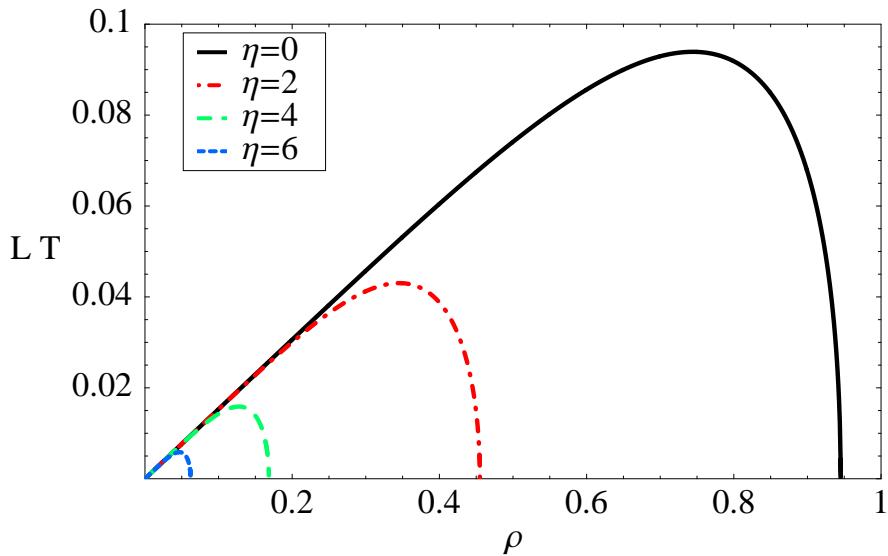}
  \caption{Baryon radius $L$ versus $\rho$, where $\rho=r_0/r_e$ is
  the ratio of the position of the black hole horizon to the position of the D5-brane,
  for several different values of the rapidity $\eta$ of the hot wind.
  The screening length $L_s$ at a given $\eta$ is the maximum
  of $L(\rho)$, namely the largest $L$ at which a static baryon
  configuration can be found.  We see that $L_s$ decreases with
  increasing wind velocity.
    }\label{fig:L_p}
\end{figure}

The integration constant $K$ must be the same at any $r$. Upon evaluating
it at $r=r_e$ and after 
some algebraic
manipulations, equations (\ref{eq:81}) and (\ref{eq:91}) can be
written more explicitly as
 \be
x' = {K \ov  \sqrt{ \le(A {r^2 \ov R^2}  - K^2 \ri) {r^2 \ov R^2}
f(r) }}\ ,
 \ee
and
\begin{equation} \label{eq:10}
\frac{K^2 R^4}{r_e^4} = 1-\rho^4 \cosh^2\eta -(1-\rho^4)\Sigma^2,
\end{equation}
from which we obtain an explicit expression for the baryon radius
$L$ in terms of $\rho$ and the rapidity $\eta$:
\begin{equation} \label{eq:11}
L = \frac{\rho}{\pi T}
\left(1-\rho^4 \cosh^2\eta
-(1-\rho^4)\Sigma^2\right)^{\frac{1}{2}} \int_1^\infty dy
\frac{1}{\left(y^4-\rho^4\right)^{\frac{1}{2}}
\left(y^4-1 +(1-\rho^4)\Sigma^2 \right)^{\frac{1}{2}} },
\end{equation}
where $y\equiv r/r_e$. We have evaluated (\ref{eq:11}) numerically, and in
Fig.~\ref{fig:L_p} we plot $L$ versus $\rho$ for several values of
$\eta$.   We see that $L$ is small when $\rho$ is small (meaning
that $r_e$ is large). As we decrease $r_e$, pulling the D5-brane
in Fig.~\ref{fig:baryon} downward, $\rho$ increases and the size
of the baryon configuration $L$ at first increases, then
reaches a maximum value, and then decreases to zero.
For a given $\eta$,  therefore, there is a maximum possible baryon
radius, which we denote $L_s$, beyond which no baryon
configurations are found. We shall identify $L_s$ with the
screening length, although in so doing we neglect a small
correction that we shall discuss below. We see from Fig.~\ref{fig:L_p}
that at any $\eta$ for $L<L_s(\eta)$ there are two
solutions with different values of $\rho$. We shall see below that
the configuration with the larger $\rho$ is unstable and has a higher energy.

According to (\ref{eq:11}), the nonzero value of $\rho$ at which $L\rightarrow 0$ 
in Fig.~\ref{fig:L_p} is the $\rho$ at which
the right-hand side of (\ref{eq:10}) vanishes. At
this value of $\rho$, $K$ is zero and $\p_r x|{r_e} = 0$,
corresponding to a configuration whose strings have become
vertical. Note that the D5-brane becomes heavier when it is closer
to the AdS black hole (i.e. $\Sigma$ in~(\ref{eq:91}) 
increases with  $\rho$), meaning that the
strings emerging from the D5-brane must be more and more vertical in order
to hold it at rest.  At some $\rho$, the strings become vertical and
at larger  $\rho$ (smaller $r_e$)
no static configuration can be found. From (\ref{eq:10}) we also
see that this largest possible $\rho$ is always smaller than  the
$\rho = 1/\sqrt{\cosh \eta}$ at which the speed $v$ exceeds
the local speed of light at the position of the D5-brane.

At any $L$ for which there are two string configurations
possible in Fig.~\ref{fig:L_p}, we expect that the solution
with the larger $\rho$ is unstable, as in the case of
the string configuration between a quark and
antiquark~\cite{Friess:2006rk}. 
This instability can be seen on qualitative grounds
as follows.  For the solutions with smaller $\rho$, we
see from Fig.~\ref{fig:L_p} that
$L$ increases monotonically with $\rho$. 
This means that if we deform the configuration by pulling the
D5-brane downward while keeping $L$ fixed, the deformed configuration
with its enlarged $\rho$ has too small an $L$ to be static. The fact that $L$
``wants'' to be larger means that the upward force on the D5-brane is
greater than required to balance the
downward force of gravity. So, there is a net restoring force pulling
the D5-brane back upwards and the original configuration is stable
against this deformation.  In contrast, for the solutions with larger
$\rho$ we see from Fig.~\ref{fig:L_p} that $L$ decreases monotonically with $\rho$,
meaning that if we pull the D5-brane downward, $L$  ``wants'' to be smaller
and the upward force on the D5-brane is less than 
the downward force of gravity (the downward force has
increased more than the upward force)
and the D5-brane will accelerate 
downward.  The configurations 
described by the part of the curve in Fig.~\ref{fig:L_p} for which $L$ decreases
with increasing $\rho$ are therefore unstable. We shall see below that
these configurations have higher energy than the stable configurations with
the same $L$ and smaller $\rho$.

We can use (\ref{eq:11}) and Fig.~\ref{fig:L_p} to 
compare the length scale
$R^2/r_e$ 
of the disturbance of the gluon field induced by the $N_c$ external
quarks to $2L$, the size of the circle of quarks itself.  In the
small-$\rho$ limit,  (\ref{eq:11}) simplifies to
\be
LT \approx \frac{0.4811\, \rho}{\pi}\ ,
\ee
which describes the linear region seen in all of the curves in Fig.~\ref{fig:L_p}
at small $\rho$. This implies that at small $\rho$
\be
\frac{R^2}{r_e} \approx 2.079\,L\ ,
\ee
comparable to $2L$.  We see from Fig.~\ref{fig:L_p} that as we go from this
small $\rho$ regime towards $L=L_s$, the ratio of $R^2/r_e$ to $2L$ increases
by a further few tens of percent.   

We see from Fig.~\ref{fig:L_p} that the screening length $L_s$
decreases with increasing velocity.  At zero velocity, $L_s=0.094/T$
as can be obtained from previous results~\cite{Brandhuber:1998xy}.
We have evaluated $L_s$ as a
function of rapidity $\eta$, and shall plot the result in
Fig.~\ref{fig:LsqrtCoshn_vs_n}, along with analogous results from
Section~\ref{sec:parallel} for the case where the wind velocity is parallel
to the plane of the baryon configuration.   From our numerical
evaluation, we find that at large $\eta$
\begin{equation} \label{eq:12}
L_s \approx\frac{a}{T \sqrt{\cosh\eta}},
\end{equation}
with $a=0.0830$.  The screening length for a quark and antiquark
separated by a distance $L^{\rm meson}$ moving through the plasma
in a direction perpendicular to the dipole also takes the form
(\ref{eq:12}) in the high velocity limit, with
$a=0.237$~\cite{Liu:2006nn}. When we compare the $L_s$ that we
have computed for the baryon configuration to $L_s^{\rm meson}/2$
(the ``radius'' of the meson configuration at its screening
length) we see that, in addition to having precisely the same
velocity dependence at high velocity, their numerical values are
comparable.  Finally, it is perhaps not surprising that $L_s^{\rm baryon}$ is somewhat 
smaller than $L_s^{\rm meson}/2$, for a given $\eta$ and $T$, since the baryon vertex
(D5-brane) pulls the strings further downward, closer to the horizon.

We can also find the large $\eta$ dependence of $L_s$
analytically. 
If we define 
\be
\hat\rho \equiv \rho \sqrt{\cosh\eta},\qquad \hat L \equiv L \sqrt{\cosh\eta}
\ee
and take the scaling limit in which 
\be
\eta\rightarrow\infty \quad {\rm with}\quad  \hat\rho,\  \hat L \quad \textrm{held fixed}, 
\label{lerl}
\ee
we find that
$\cosh \eta$ drops out of the leading terms in Eq.~(\ref{eq:11}) and this
equation becomes 
\begin{eqnarray} \label{eq:14}
\hat L & = &
 \frac{\hat \rho}{\pi T}
\left(1-\hat \rho^4  - \Sigma^2\right)^{\frac{1}{2}} \int_1^\infty
dy \frac{1}{y^2 \left(y^4 - 1+\Sigma^2 \right)^{\frac{1}{2}}
} + O\le((\cosh \eta)^{-\ha} \ri) \:\:\:\:\: \nonumber\\
 &=&\frac{\hat \rho}{3 \pi T}
\left(1-\hat \rho^4 - \Sigma^2\right)^{\frac{1}{2}}
\:{_2}F_1\left(\frac{1}{2},\frac{3}{4},\frac{7}{4}, 1-\Sigma^2
\right) + O\le((\cosh \eta)^{-\ha} \ri) \ .
\end{eqnarray}
(Note that according to (\ref{eq:91}), $\Sig$ only
depends on $\hat \rho$.)
The right-hand side of (\ref{eq:14}) is function of $\hat \rho$
that goes to zero at $\hat \rho\rightarrow 0$ and 
at $\hat\rho\rightarrow 0.880$, and that has a maximum at
$\hat\rho = 0.666$ where $\hat L =0.0830/T$, yielding an
$L_s$ that is in precise agreement with~(\ref{eq:12}).

We close this section by evaluating the energy of the baryon
configurations that we have constructed. The energy of one string
can be found using $S_{\textrm{string}}$ and is given by
\begin{eqnarray} \label{eq:16}
E_{\mathrm{string}} &=& \frac{1}{2\pi \alpha'}\int_{r_e}^{\infty}
dr \sqrt{A\left(\frac{(x')^2 \,r^2}{R^2}+\frac{1}{f}\right)}\nonumber\\ &=& \frac{T\sqrt{\lambda}}{2
\rho}\int_1^\infty dy \frac{y^4-\rho^4
\cosh^2\eta}{\left(y^4-\rho^4\right)^{\frac{1}{2}}\left(y^4-1+(1-\rho^4)\Sigma^2\right)^{\frac{1}{2}}}\
,
\end{eqnarray}
where $y\equiv r/r_e$. This energy is infinite because we have
included the masses of the quarks.  As in
Refs.~\cite{Liu:2006nn,Liu:2006he}, we regularize the baryon
energy by subtracting the energy of (in this case $N_c$) disjoint
quarks in a hot plasma wind of velocity $v$, whose strings are dragging behind
them in the $x_3$ direction according to the solution found
in~\cite{Herzog:2006gh,Gubser:2006bz}. This corresponds to
regulating the $r$-integral in (\ref{eq:16}) with a large-$r$
cutoff $\Lambda$, subtracting
\begin{equation} \label{eq:17}
E_{\mathrm{mass}} = \frac{N_c}{2\pi\alpha'} \int_{r_0}^\Lambda dr
= \frac{N_c T \sqrt{\lambda} }{2\rho}\int_\rho^{\Lambda/r_e} dy\ ,
\end{equation}
and then taking $\Lambda$ to infinity. This procedure yields a
finite answer. The total energy of the baryon (strings plus
D5-brane) becomes
\begin{eqnarray} \label{eq:18}
E_{\mathrm{total}}&=&\frac{N_c T\sqrt{\lambda}}{2} \Biggl[
\frac{1}{\rho}\int_1^\infty dy \left(\frac{y^4-\rho^4
\cosh^2\eta}{\left(y^4-\rho^4\right)^{\frac{1}{2}}
\left(y^4-1+(1-\rho^4)\Sigma^2\right)^{\frac{1}{2}}}-1\right) + 1-
\frac{1}{\rho}
\nonumber \\
&~&\qquad\qquad + \frac{\sqrt{1-\rho^4\cosh^2\eta}}{4
\rho}\Biggr]\ ,
\end{eqnarray}
where the last term is the contribution of the D5 brane to the
energy.

\begin{figure}
  \centering
  \includegraphics*[width=0.8\columnwidth]{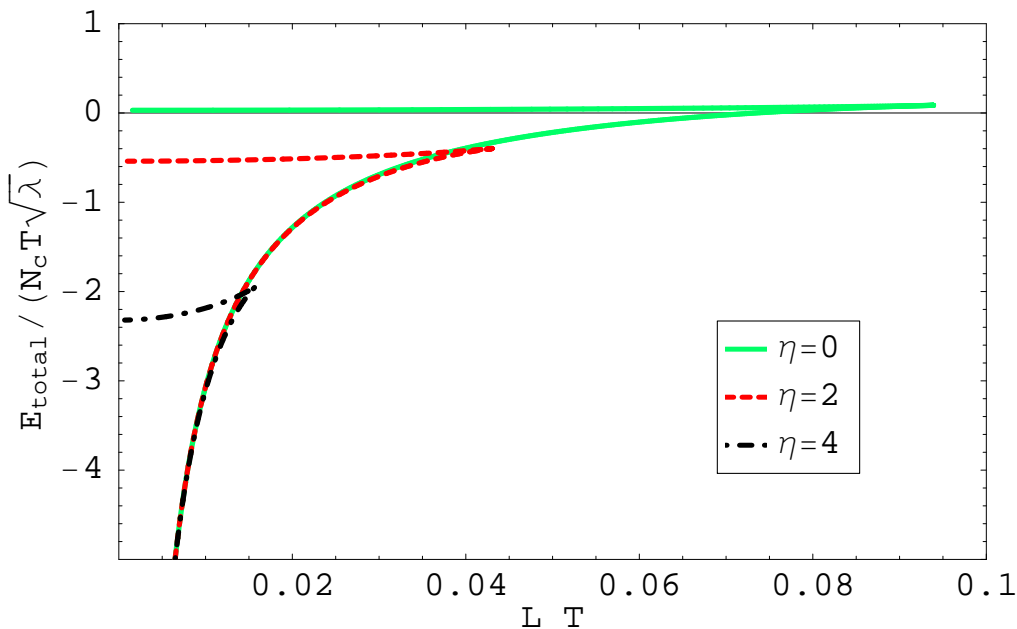}
  \caption{The total energy of the baryon configuration
  with a given $L$ (relative
  to that of $N_c$ disjoint quarks moving with the
  same velocity) for several values of the wind rapidity $\eta$.
The lower (higher) energy branch corresponds to the solution in
Fig.~{\protect\ref{fig:L_p}} with lower (higher) $\rho$.  The
cusps where the two branches meet are at $L=L_s$.
    }
  \label{fig:E_L}
\end{figure}

In Fig.~\ref{fig:E_L} we plot the energy of the baryon
configurations at several values of $\eta$.  As in
Fig.~\ref{fig:L_p}, we see two configurations at every $L<L_s$.
We have argued above that
the higher energy configurations (those with the larger $\rho$)
are unstable, so we focus on the lower branch. We see
that at $\eta=0$ this branch crosses zero energy at $L=0.073/T$,
whereas the largest $L$ at which a baryon configuration can be
found is $L_s=0.094/T$. This means that for $0.073<LT<0.094$, even
the lower branch solutions have become metastable, as they have
more energy than $N_c$ disjoint quarks.   We see from
Fig.~\ref{fig:E_L} that this phenomenon does not occur at larger
velocities; in fact, it arises only for $\eta\leq 0.755$ since at
$\eta=0.755$ the baryon configuration with $L=L_s$ has the same
energy as $N_c$ disjoint quarks, i.e. zero energy in
Fig.~\ref{fig:E_L}. At the low velocities $\eta<0.755$, a more
precise definition of the screening length would be to define it
as the length at which the lower curve in Fig.~\ref{fig:E_L}
crosses zero.  We see from Fig.~\ref{fig:E_L} that by using $L_s$
as our definition of the screening length at all velocities, as we
do for simplicity, we are introducing only a small imprecision at
low velocities, $\eta<0.755$.  These considerations have no effect
on our calculation of $L_s$ at large $\eta$, namely (\ref{eq:12}).

It is clear from (\ref{eq:18}) that in the large $\eta$
limit~(\ref{lerl}) with $\hat L$ held fixed and $L$ therefore
decreasing, the energy scales as $-\sqrt{\cosh \eta}$.
This scaling can also be deduced from Fig.~\ref{fig:E_L}, as follows.
The subtraction term (\ref{eq:17}) is defined such
that for any given $T$ and $\eta$, at small enough $L$ the
potential energy $E(L)$ is the same as in vacuum (i.e. for $T=\eta=0$),
namely $E(L)\propto -1/L$. 
And, if $E\propto-1/L$  and the $E(L)$ curves for different $\eta$ overlap
as seen in Fig.~\ref{fig:E_L}, then 
$E$ must scale like $-\sqrt{\cosh\eta}$
in the limit (\ref{lerl}).

The baryon configuration that we have analyzed in this 
subsection is special in that all $N_c$ strings are equivalent.
In the next subsection we shall analyze a configuration for
which this is not the case, for which we shall need the
full formalism developed in Section 2.

\subsection{Wind parallel to the baryon
configuration}\label{sec:parallel}

We now analyze the case where the $N_c$ quarks are moving through
the plasma (or, equivalently in their rest frame, feeling a hot
wind blowing) in a direction parallel to their plane.  
We shall keep the wind blowing in the $x_3$ direction as before, 
meaning that the
boosted AdS black hole metric given by (\ref{eq:1})
is unchanged. We shall now
put the $N_c$ quarks in the $(x_1,x_3)$-plane. 

With the quarks in
the $(x_1,x_3)$-plane and the wind velocity in the $x_3$
direction, the $N_c$ strings in a circular baryon configuration
are no longer equivalent,
as the strings make different angles relative
to the wind direction.  The $N_c$ strings would not all hit the
D5-brane at the same angle in this case.  
Analyzing this case is possible, but
we will instead consider a simpler configuration in which all
$N_c$ strings hit the D5-brane symmetrically. In terms of the
formalism developed in Section~\ref{sec:gen}, we 
choose a configuration in which the string
density at the D5-brane is
 \be \label{ep}
 \rho_V (s_1, s_2,s_3) = {1 \ov  \pi} \delta (s_1^2 + s_3^2 -s^2)\, \delta(s_2)\ ,
 \ee
where $s$ is some constant and $s_i = \p_r x_i (r)|_{r_e}$. 
The distribution (\ref{ep}) corresponds to requiring that the
$N_c$ strings hit the D5-brane with a uniform distribution
in the azimuthal angle $\phi$ in the $(x_1,x_3)$-plane and all 
with the same
$ \p_r x|_{r_e}=s$. (Here, $x\equiv \sqrt{x_1^2+x_3^2}$.)
Unlike in the previous section, this specification of the baryon
configuration in the vicinity of $r=r_e$ will {\it not} correspond to
having the $N_c$ quarks at $r=\infty$ arranged on a circle.
Note that (\ref{ep}) guarantees that the net force exerted on the
D5-brane in the $x_1$- and $x_3$-directions by the $N_c$ quarks
vanishes, meaning that (\ref{Bss3}), or equivalently (\ref{Bss}), is  
automatically satisfied.
Given the choices that we have made in specifying our
baryon configuration, our task is twofold: we must determine $s$
as a function of $r_e$
such that the forces on the D5-brane in the
$r$-direction due to gravity and due to the strings cancel; and,
we must solve for the shape of the strings to determine what
baryon configuration at $r=\infty$ our choices correspond to.

The shape of each string is specified by two functions $x_1 (r)$
and $x_3 (r)$ that we must obtain. We shall find that, when
projected onto the $(x_1,x_3)$-plane, the string worldsheets do
not follow radial trajectories.  That is, the trajectories
$x_1(r)$ and $x_3(r)$ are not specified just by $x(r)$;  their
azimuthal angle $\phi$ is a nontrivial function of $r$ also.

Applying equations (\ref{eq:6}) and (\ref{d5A}) to (\ref{eq:1})
with nontrivial $x_1 (r)$ and $x_3 (r)$, we find that 
\begin{equation} \label{eq:5b}
{\cal L}_{\rm string} = \sqrt{A
\left(\frac{1}{f(r)}+\frac{(x'_1)^2\, r^2}{R^2} \right)+
\frac{(x'_3)^2 \, r^2 f(r)}{R^2}} \ ,
\end{equation}
and find that the D5-brane action is given by~(\ref{eq:S_D5}) as before. With
$\mathcal{L}_{\textrm{string}}$ given by~(\ref{eq:5b}), the
equations of motion (\ref{eq:8}) can be rearranged to give
\begin{equation} \label{eq:9b}
x_1'^2=\left(\frac{R^4}{r^2}\right)\left(\frac{K_1^2
}{r^2 f A - R^2 K_3^2 A - R^2 K_1^2 f}\right),
\end{equation}
\begin{equation} \label{eq:9c}
x_3'^2=\left(\frac{R^4}{f^2r^2}\right)\left(\frac{K_3^2 A^2}{r^2 f
A -R^2 K_3^2 A - R^2 K_1^2 f}\right).
\end{equation}
Equation (\ref{vre}) for the balance of force in the radial
direction becomes
 \begin{equation} \label{eq:14b}
\sum_{\mathrm{strings}} \frac{R \,A/\sqrt{f}}{\sqrt{A \left(R^2 +
f r^2 x_1'^2 \right)+ f^2 r^2 x_3'^2}}
\biggr|_{r_e}=N_c \Sigma(\rho,\eta)\ ,
\end{equation}
where $\Sigma(\rho,\eta)$ is as in (\ref{eq:91}) and is the downward
gravitational force on the D5-brane and the left-hand side of (\ref{eq:14b}) is
the upward force due to the $N_c$ strings.
If we define $\phi$ to be the azimuthal angle in the $(x_1,x_3)$-plane 
that a string makes at $r=r_e$ where it connects to the
D5-brane, defined such that $\phi=0$ ($\phi=\pi/2$) is in the
positive-$x_3$ (positive-$x_1$) direction,
then our choice of having the $N_c$ strings uniformly distributed
in $\phi$ turns the sum over strings in equation (\ref{eq:14b})
into an integral over $\phi$,
\begin{equation} \label{eq:17b}
\sum_{\mathrm{strings}}\rightarrow N_c \int_0^{2 \pi}
\frac{d\phi}{2\pi},
\end{equation}
and expression (\ref{eq:14b}) becomes
\begin{equation} \label{eq:19b}
\frac{R\,A}{\sqrt{f}}\int_0^{2\pi} \frac{d\phi}{2\pi}
\frac{1}{\sqrt{A R^2+s^2 f r^2 \left(A \sin^2\phi+f
\cos^2\phi\right)}}\Biggr|_{r_e}=\Sigma(\rho,\eta),
\end{equation}
where  $s=\partial_r x|_{r_e}$ was introduced in (\ref{ep}) and as before $\rho\equiv
r_0/r_e$.

The constants $K_1$ and $K_3$ must be the same at any $r$. By
evaluating (\ref{eq:9b}) and (\ref{eq:9c}) at $r=r_e$ and
rearranging, we determine that
\begin{equation} \label{eq:K1}
K_1^2=\frac{s^2 A^2 r^4 f \sin^2\phi}{R^2 \left(A R^2+s^2 r^2 f
\left(A \sin^2\phi+f \cos^2\phi\right)\right)}\biggr|_{r_e},
\end{equation}
\begin{equation} \label{eq:K3}
K_3^2=\frac{s^2 r^4 f^{3} \cos^2\phi}{R^2 \left(A R^2+s^2 r^2
f\left(A \sin^2\phi+f \cos^2\phi\right)\right)} \biggr|_{r_e} \ .
\end{equation}
With these integration constants now determined, we can integrate
Eqs.~(\ref{eq:9b}) and (\ref{eq:9c}),  obtaining
\begin{equation} \label{eq:23b}
x_1(r)= \frac{\rho^3 R^4 K_1}{r_0^3}\int_1^{r/r_e} d y
\frac{1}{\sqrt{Q}}\ ,
\end{equation}
and
\begin{equation} \label{eq:24b}
x_3(r)= \frac{\rho^3 R^4 K_3}{r_0^3}\int_1^{r/r_e} d y\:\:
\frac{y^4-\rho^4 \cosh^2\eta}{y^4-\rho^4}\frac{1}{\sqrt{Q}}\ ,
\end{equation}
where
\begin{equation} \label{eq:24c}
Q\equiv(y^4-\rho^4)(y^4-\rho^4\cosh^2\eta)-\frac{R^4\rho^4
K_1^2}{r_0^4}(y^4-\rho^4)-\frac{R^4\rho^4
K_3^2}{r_0^4}(y^4-\rho^4\cosh^2\eta).
\end{equation}
Equations (\ref{eq:23b}) and (\ref{eq:24b}) specify the shape of
the string worldsheets, while $r_e$ (equivalently, $\rho$) is
determined in terms of $s$ by (\ref{eq:19b}).

\begin{figure}
  \centering
  \includegraphics*[width=0.7\columnwidth]{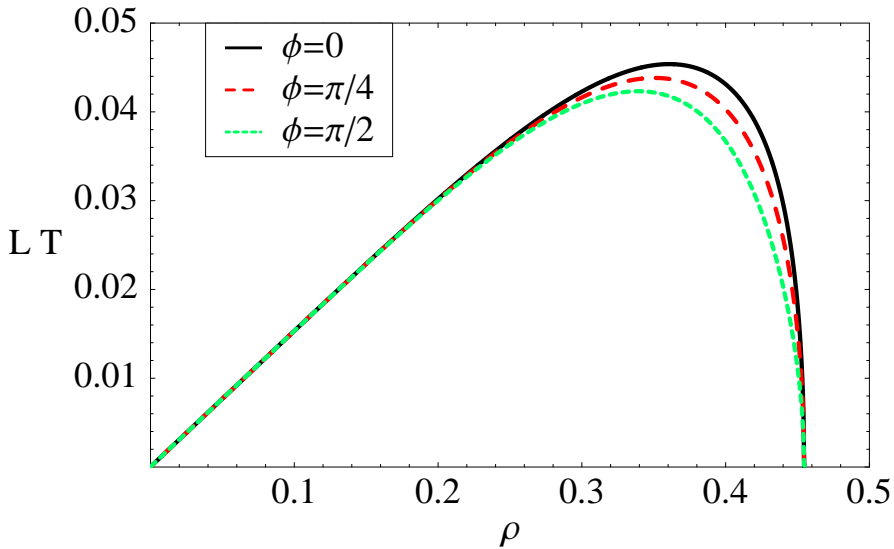}
  \caption{$L$ versus $\rho$ for strings oriented in the $\phi=0,\pi/4,\pi/2$
  directions in the $(x_1,x_3)$-plane in a baryon configuration immersed
  in a plasma moving in the $x_3$-direction with rapidity $\eta=2$.
  We see that at a given $\rho$ the distance $L$ in the $(x_1,x_3)$-plane
  between a quark and
  the D5-brane at the center of the baryon configuration depends
  on the angular position of the quark.  This means that the $N_c$ quarks
  do not lie on a circle.
    }\label{fig:L_p_horizontal}
\end{figure}
\begin{figure}
  \centering
\includegraphics*[width=0.59\columnwidth]{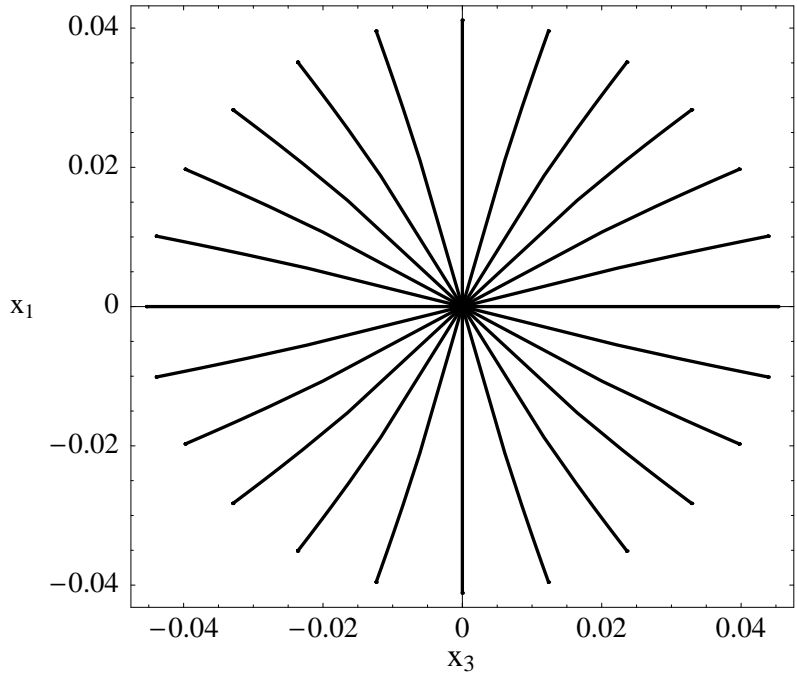}
  \caption{Strings projected on the AdS boundary
for $\eta=2$ and $\rho=0.37$ for strings separated in $\phi$ by
$\pi/12$. (We have done all our calculations for
$N_c\rightarrow\infty$, but have shown only 24 quarks in the
figure.) Baryon motion is in the $x_3$ direction. The figure is
drawn in the rest frame of the baryon, meaning there is a hot wind
in the $x_3$ direction. The $N_c$ quarks that make up the baryon
configuration are not arranged in a circle: the ``squashed circle"
is wider in the direction of motion. Note also that the projection
of the strings are not straight lines.}
  \label{fig:x1_x3_n2_p37}
\end{figure}
\begin{figure}[h!]
  \centering
\includegraphics*[width=0.59\columnwidth]{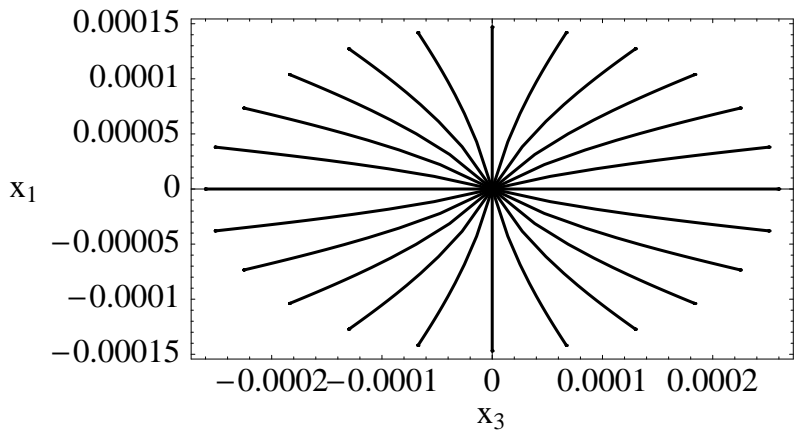}
  \caption{Same as Fig.~{\protect\ref{fig:x1_x3_n2_p37}}, but for
  $\rho=0.4550611$, very close to the maximum $\rho$ at which a
  baryon configuration can be found at $\eta=2$.
  This configuration is unstable, and 
  has higher energy than the configurations with comparable
  $L$'s at much lower $\rho$. However, it illustrates the ``squashing'' of the
  baryon configuration away from a circular shape and the curvature of the
  projections of the strings onto the $(x_1,x_3)$ plane. Both these effects
  are present in  Fig.~{\protect\ref{fig:x1_x3_n2_p37}},  but are more
  visible here. 
  \vspace{-0.4in}
}
  \label{fig:x1_x3_n2_p4550611}
\end{figure}

The calculation proceeds as follows. First, we solve
(\ref{eq:19b}) numerically to obtain the $s$ required at a given
$\rho$. Second, we pick a particular value of $\phi$ and use $s$
to evaluate (\ref{eq:K1}) and (\ref{eq:K3}), obtaining the
$r$-independent, but $\phi$-dependent, $K_1$ and $K_3$.    Third,
we evaluate (\ref{eq:23b}) and (\ref{eq:24b}) numerically, thus
obtaining the shape of the string with a particular value of
$\phi$. The position of the quark at $r=\infty$ corresponding to
this choice of $\phi$ is then $(x_1(\infty),x_3(\infty))$ and we
can determine $L=\sqrt{x_1(\infty)^2+x_3(\infty)^2}$ for this
choice of $\phi$. Fourth, we repeat the second and third steps for
all values of $\phi$.

In Fig.~\ref{fig:L_p_horizontal} we show the $L$ obtained as we
have just described at three values of $\phi$, as a function of
$\rho$. We conclude from the fact that $L$ is different for
different values of $\phi$ that the $N_c$ quarks at $r=\infty$ are
not arranged in a circle in the $(x_1,x_3)$-plane. We illustrate
this explicitly in Figs.~\ref{fig:x1_x3_n2_p37} and
\ref{fig:x1_x3_n2_p4550611}. We started with circularly symmetric
boundary conditions at the D5-brane, but the resulting baryon
configuration at the AdS boundary is ``squashed", wider in the
direction of motion of the baryon and narrower in the
perpendicular direction. Inspection of Fig.~\ref{fig:L_p_horizontal}
or comparison of Figs.~\ref{fig:x1_x3_n2_p37} and
\ref{fig:x1_x3_n2_p4550611} shows that the shape of the
baryon configuration at the AdS boundary changes with $\rho$,
becoming more squashed as $\rho$ increases.
In Figs.~\ref{fig:x1_x3_n2_p37} and \ref{fig:x1_x3_n2_p4550611} 
we also see that the projections of the
string worldsheets onto the $(x_1,x_3)$-plane are not straight
radial lines.  Their  curved shapes are strikingly similar to the
shapes of the projections of strings joining a static
quark-antiquark in the meson configurations analyzed in
Refs.~\cite{Liu:2006nn,Liu:2006he}, although they are not
precisely the same. Note that Eqs.~(\ref{eq:19b})--(\ref{eq:K3}) 
are symmetric in $\phi \rightarrow \pi - \phi$, which implies that 
string configurations are
symmetric with respect to 
reflection in the $x_1$ axis, i.e. under $x_3 \rightarrow
-x_3$, as is manifest in Figs.~\ref{fig:x1_x3_n2_p37} and
\ref{fig:x1_x3_n2_p4550611}.
 This forward-backward symmetry
 of the string configurations  indicates that the baryon configuration  feels no
drag as it is moved through the plasma, just as for meson
configurations~\cite{Peeters:2006iu,Liu:2006nn,Liu:2006he}, and as has
been demonstrated previously for baryon configurations with
zero size \cite{Chernicoff:2007}.

It is straightforward to compute the energy of the baryon configurations
that we have found, as a function of $\rho$, but since 
(unlike in Section~\ref{sec:perpend}) the configurations
are not characterized by a single $L(\rho)$ there is no analogue
of Fig.~\ref{fig:E_L} here.  Also, 
(again unlike when the  wind blows
perpendicular to the baryon configuration as in Section~\ref{sec:perpend})
we have no simple argument for at what $\rho$ the baryon configurations
in this section become unstable.  Our argument in the previous section
relied on the equivalence of all $N_c$ strings, in that at a single $\rho$ 
there was a change from ``a deformation that increases $\rho$ makes
all $N_c$ strings want to have larger $L$'' to ``a deformation that increases
$\rho$ makes all $N_c$ strings want to have smaller $L$".  Here, 
we see from Fig.~\ref{fig:L_p_horizontal} that there
is a range of $\rho$ within which a deformation that increases $\rho$
makes some strings want to have smaller $L$ while other strings
want to have larger $L$. Within this range of $\rho$, our simple argument
yields no conclusion and a full stability analysis as in Refs.~\cite{Friess:2006rk} is
required. We leave this to future work.

\begin{figure}
  \centering
  \includegraphics*[width=0.8\columnwidth]{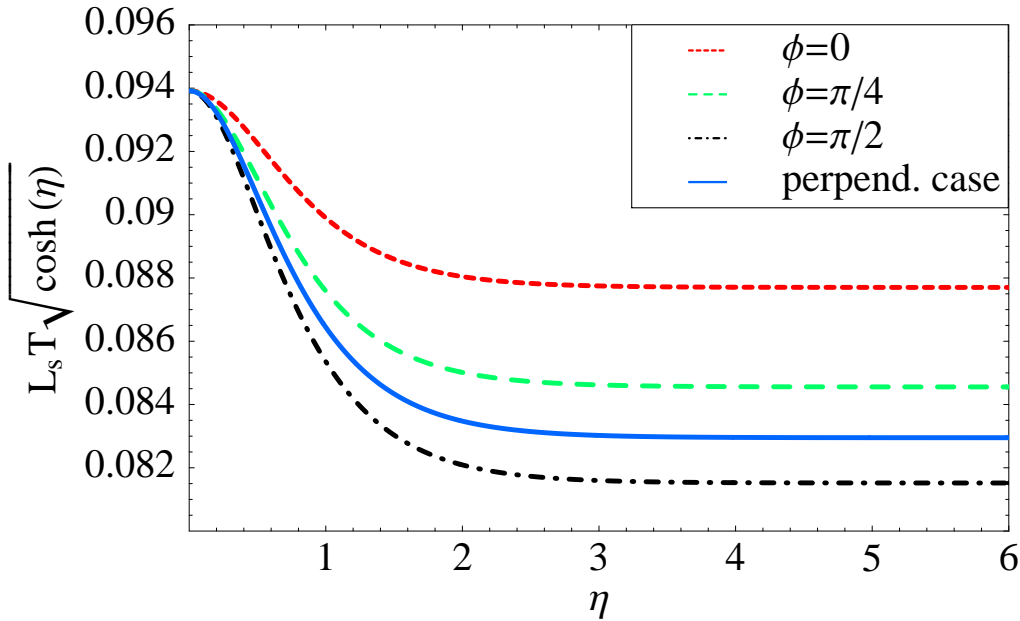}
  \caption{The screening length $L_s$ as a function of $\eta$ with its large-$\eta$ dependence
  $\sqrt{\cosh\eta}$ scaled out.  The solid curve is for the case of a wind velocity perpendicular
  to the plane of the baryon, as in Section~{\protect\ref{sec:perpend}}.  The other three curves
  are for wind velocity in the plane of the baryon, and show the $L_s$
  for the strings  that make an angle $\phi=0,\pi/4,\pi/2$ with the direction of the wind.
    }\label{fig:LsqrtCoshn_vs_n}
\end{figure}

\begin{figure}
  \centering
  \includegraphics*[width=0.7\columnwidth]{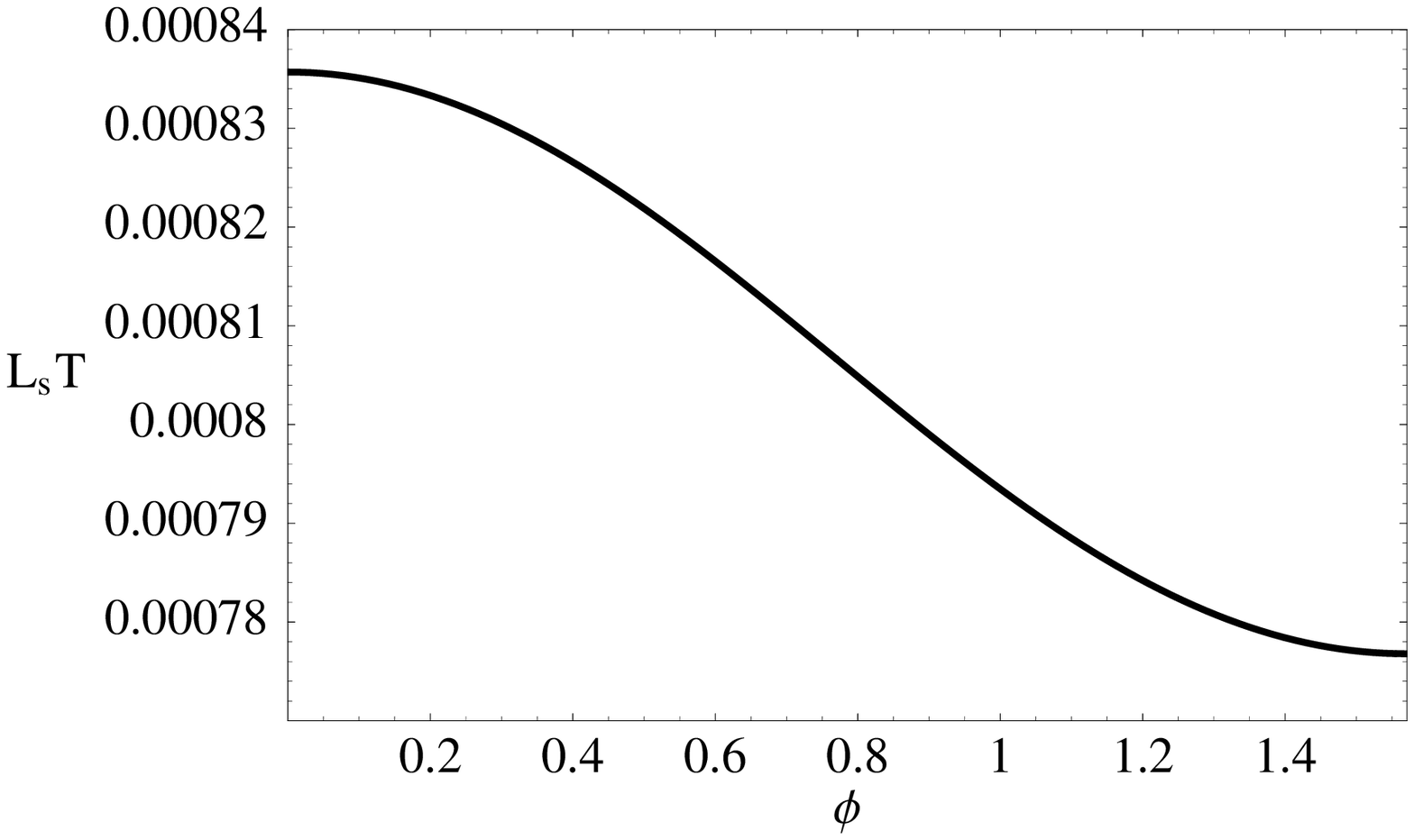}
  \caption{The screening length $L_s$ as a function of $\phi$ at a large value of $\eta$. Specifically,
  $\eta=10$ corresponding to $\sqrt{\cosh \eta } = 1/(1-v^2)^{1/4} = 104.9$.
    }\label{fig:L_phi}
\end{figure}

The maxima of curves as in Fig.~\ref{fig:L_p_horizontal} define a
screening length $L_s$ for each $\phi$ as a function of $\eta$.
In Fig.~\ref{fig:LsqrtCoshn_vs_n} we plot $L_s T\sqrt{\cosh\eta} =
L_s T/(1-v^2)^{1/4}$ versus $\eta$ for different values of $\phi$.
We find that the large-$\eta$ dependence of the screening length
has the same form in all cases, namely
\begin{equation} \label{eq:27b}
L_{s,\:\eta\gg1}\propto \frac{1}{T\sqrt{\cosh\eta}}\ .
\end{equation}
This is the same large $\eta$ dependence found in
Section~\ref{sec:perpend}, Eq.~(\ref{eq:12}), and in mesons,
Eq.~(\ref{lmax}).  To make the former comparison manifest, in
Fig.~\ref{fig:LsqrtCoshn_vs_n} we have also plotted $L_s T
\sqrt{\cosh\eta}$ for the case analyzed in
Section~\ref{sec:perpend} in which the wind velocity is
perpendicular to the plane of the baryon configuration. When the
wind velocity is parallel to the plane of the baryon
configuration, $L_s$ has a weak angular dependence. In particular,
the constant of proportionality in Eq.~(\ref{eq:27b}) varies
between $0.082$ and $0.088$ for different $\phi$, as can be seen
in Fig.~\ref{fig:LsqrtCoshn_vs_n}. A plot of $L_s$ in the large
$\eta$ regime as a function of $\phi$ is given in
Fig.~\ref{fig:L_phi}, which shows the smooth variation of $L_s$
for large $\eta$ as we change $\phi$. Note also that
(\ref{eq:27b}) is a good approximation all the way down to the
small velocity limit $\eta\rightarrow 0$, since the
proportionality constant in Eq.~(\ref{eq:27b}) merely changes from
its ($\phi$-dependent) value at large $\eta$  to  the (obviously
$\phi$-independent) value $0.094$ at $\eta=0$. The central
conclusion to be drawn from Fig.~\ref{fig:LsqrtCoshn_vs_n}, then,
is that the simple velocity scaling (\ref{SimpleScaling})  is a
good approximation at all velocities and all angles.

The similarities between our results and those for the meson
screening length go beyond just the dominant velocity scaling.
Indeed, Fig.~\ref{fig:LsqrtCoshn_vs_n} is strikingly similar to
Fig.~7 of Ref.~\cite{Liu:2006he}.  There too, for the
quark-antiquark case, $L_s T \sqrt{\cosh\eta}$ is largest at
$\eta=0$, a few percent smaller for $\eta\rightarrow\infty$ if the
quark-antiquark dipole is oriented parallel to the wind, and a few
percent smaller still if the dipole is oriented perpendicular to
the wind.  The only feature in our Fig.~\ref{fig:LsqrtCoshn_vs_n}
that does not have a direct, almost quantitative, analogue in
Ref.~\cite{Liu:2006he} is the {\it very} small difference between
the curves for the two cases in which the wind direction is
perpendicular to the line between the quark and the D5-brane,
namely the case in which the wind is parallel to the plane of the
baryon configuration and the quark is at $\phi=\pi/2$ and the case
in which the wind is perpendicular to the plane of the baryon
configuration.

Although we have only done our analysis for a wind that is either
perpendicular to or parallel to the plane of the baryon
configuration, we expect that the qualitative features that we
have found in this section should all be present for any wind
direction except perpendicular.

In Subsections \ref{sec:perpend} and \ref{sec:parallel} we have
analyzed two particular baryon configurations that suffice to make
our point regarding the velocity dependence of baryon screening.
The general formalism of Section~\ref{sec:gen} can straightforwardly be
applied to baryon configurations with 
other shapes, whether specified by 
the density of quarks at infinity or the density of strings at the D5-brane vertex.
Technically, in order to solve equations (\ref{Bss}) and (\ref{vre}), it is
simpler to specify the density of strings at the D5-brane as we have
done in this subsection, but there is no obstacle of principle to 
analyzing arbitrary densities of quarks at infinity in any wind velocity.
While the behavior at small $\eta$ could
differ for more general configurations, we expect that in the large $\eta$ limit, the
scaling behavior (\ref{eq:27b}) should still apply.
The formalism of Section~\ref{sec:gen} can also be used  to generalize
our results to the plasmas of other strongly coupled gauge theories.
For example,  following a line
of reasoning developed in Ref.~\cite{Caceres:2006ta} for the meson sector,
it can be shown that in a certain class of gauge theories whose gravity duals are
asymptotically AdS, as $v\rightarrow 1$ the baryon screening length scales as 
$L_s \propto (1-v^2)^{1/4}/\eps^{1/4}$, 
where $\eps$ is the energy density of
the plasma. $\eps$  is proportional to $T^4$ for the specific
theory that we have analyzed, at any $v$, in this Section.

\section{Discussion}\label{sec:conclusions}

We have analyzed the screening of the static potential for a
baryon configuration consisting of $N_c$ quarks in a circle (or
slightly squashed circle) moving with velocity $v$ through the
plasma of ${\cal N}=4$ SYM theory in a direction perpendicular (or
parallel) to the plane of the configuration.  We find a screening
length 
\be L_s = \frac{a (1-v^2)^{1/4} }{T}\ ,
\label{ConclusionEquation} 
\ee 
where $a$ depends only weakly on
$v$ and angles.  For example, $a=0.094$ for $v=0$ while
$a=0.083$ for $v\rightarrow 1$ with the direction of motion
perpendicular to the plane of the baryon configuration, and
$0.082<a<0.088$ for the case where the motion is parallel to the
plane, again for $v\rightarrow 1$.  In this last case, $a$ is
smallest for those quarks on the circle which are connected to the
D5-brane junction at the center of the baryon by a string that is
perpendicular to the direction of motion. The velocity dependence
in (\ref{ConclusionEquation}) is precisely the same as that for
the screening length defined by a quark and antiquark moving
through the plasma, and even the weak angular dependence of $a$ is
strikingly similar.  This is a confirmation of the robustness of
the velocity dependence of screening  that in the meson sector has
as a consequence the experimentally testable prediction that in a
range of temperatures that is plausibly accessed in heavy ion
collisions at RHIC (or at the LHC) $J/\Psi$ (or $\Upsilon$)
suppression may set in only for quarkonia moving with a transverse
momentum above some threshold~\cite{Liu:2006nn}. In the baryon sector, it 
suggests
that if baryons composed of three charm quarks are ever studied in
heavy ion collision experiments which do not reach such high
temperatures as to dissociate them at rest,  their production
could also be suppressed above some threshold transverse
momentum. 

We have found that if the baryon configuration we study feels
a wind velocity parallel to its plane (and presumably at any
direction except perpendicular) the $N_c$ quarks are not all
equivalent.  Those in a direction perpendicular to the wind
are most affected by the wind velocity, as in the configuration
we analyze with azimuthally symmetric boundary conditions
at the D5-brane they are the ones that are pulled in closest to
the D5-brane and yet they are also the ones that have the smallest $L_s$.
It is tempting to conclude from this that as a function of increasing
$v$ or $T$ these quarks will dissociate first. However, justifying such a
conclusion requires further work.  It could be interesting to
investigate configurations that are held circular 
in a wind parallel to their plane, which would no longer have azimuthally
symmetric boundary conditions at the D5-brane. This would allow
the analysis of a sequence of configurations with the same shape but
different size rather than a sequence of configurations whose degree
of squashing changes with size, as in our analysis. However, a definitive
conclusion
requires comparing the energies of a baryon configuration
on the one hand with a well-separated quark and $(N_c-1)$-quarks$+$D5-brane
configuration, each trailing a dragging string, on the other hand.   If  the
varying effectiveness of the screening of the potential binding different
quarks to the baryon that we have found 
were to manifest itself as some quarks dissociating
before others,  as a function of increasing $T$ or $v$, 
this would suggest that when heavy baryons
with $N_c=3$ dissociate while moving through
a strongly coupled plasma, they may at
least initially dissociate into a quark and a diquark.

\medskip
\section*{Acknowledgments}
\medskip

We acknowledge helpful conversations with Wit Busza,
Qudsia Ejaz, Tom Faulkner, Rob Myers, Gunther Roland, Aninda Sinha and 
Urs Wiedemann. 
HL is supported in part by the A.~P.~Sloan Foundation and the U.S.
Department of Energy (DOE) OJI program.  This research was
supported in part by the DOE Offices of Nuclear and High Energy
Physics under grants \#DE-FG02-94ER40818 and \#DE-FG02-05ER41360.

\end{document}